\documentclass[aps,preprint,twocolumn,10pt,superscriptaddress]{revtex4}%
\usepackage{amsfonts}
\usepackage{amsmath}
\usepackage{amssymb}
\usepackage{graphicx}
\usepackage{dcolumn}%
\begin{document}
\title{Consistent model of magnetism in ferropnictides}
\author{A. L. Wysocki}
\affiliation{Department of Physics and Astronomy and Nebraska Center for Materials and
Nanoscience, University of Nebraska-Lincoln, Lincoln, Nebraska 68588, USA}
\author{K. D. Belashchenko}
\affiliation{Department of Physics and Astronomy and Nebraska Center for Materials and
Nanoscience, University of Nebraska-Lincoln, Lincoln, Nebraska 68588, USA}
\author{V. P. Antropov}
\affiliation{Ames Laboratory, Ames, Iowa, 50011, USA}
\keywords{ferropnictides, biquadratic interaction, spin wave spectrum, phase
transitions, nematic phase, CaFe$_{2}$As$_{2}$}

\begin{abstract}
The discovery of superconductivity in LaFeAsO introduced the ferropnictides
as a major new class of superconducting compounds with critical temperatures
second only to cuprates. The presence of magnetic iron makes ferropnictides
radically different from cuprates. Antiferromagnetism of the parent compounds
strongly suggests that superconductivity and magnetism are closely related.
However, the character of magnetic interactions and spin fluctuations in
ferropnictides, in spite of vigorous efforts, has until now resisted
understanding within any conventional model of magnetism. Here we show that
the most puzzling features can be naturally reconciled within a rather simple
effective spin model with biquadratic interactions, which is consistent with
electronic structure calculations. By going beyond the
Heisenberg model, this description explains numerous experimentally observed
properties, including the peculiarities of the spin wave spectrum, thin
domain walls, crossover from first to second order phase transition under
doping in some compounds, and offers new insight in the occurrence of the
nematic phase above the antiferromagnetic phase transition.

\end{abstract}
\date{\today}
\maketitle

\section{Introduction}

The recent discovery of superconductivity in ferropnictides has ended the
hegemony of cuprates and ushered in the new ``iron era'' in the studies of
high-temperature superconductivity \cite{Mazin,Lynn,Johnston}. All parent
ferropnictide compounds, unlike cuprates, are metallic and contain iron as
the indispensable element. The occurrence of antiferromagnetism in the parent
compounds of both cuprates and ferropnictides suggests that magnetism and
superconductivity may be interrelated, and that the pairing mechanisms
involve spin fluctuations in both cuprates and ferropnictides. However,
magnetism in ferropnictides demonstrates a number of unique and puzzling
properties, which are inconsistent with any conventional model. Understanding
the character of magnetic interactions and spin fluctuations in
ferropnictides is therefore of paramount importance for the problem of
high-temperature superconductivity. In particular, a reasonable effective
spin Hamiltonian supported by experimental measurements and theoretical
calculations is highly desirable. Until now, such spin Hamiltonian has not
been found. The purpose of this article is to solve this problem.

The antiferromagnetic (AFM) stripe ground state may be obtained in the
Heisenberg model with appropriately chosen exchange parameters, such as the
$J_{1}$-$J_{2}$ model with $J_{2}>J_{1}/2$, which has been extensively
studied in the context of frustrated magnetism \cite{Larkin}. However, this
model neglects the large anisotropy of the nearest-neighbor exchange
constant, which is firmly established both from first-principles calculations
\cite{Yin} and by neutron scattering measurements
\cite{Lynn,AMESneut,ORNLneut}. The three-dimensional character of magnetic
interactions is also important \cite{Diallo}. Without these features, the
Heisenberg model can not reproduce the qualitative features of the spin-wave
spectrum. It is common to introduce the anisotropy of the exchange parameters
explicitly; in particular, $J_{1}$ is split into $J_{1a}$ and $J_{1b}$ for
the pairs with antiparallel and parallel spins, respectively. Theoretically
these parameters can be calculated using the linear response (LR) technique.
The anisotropy of $J_{1}$ strongly depends on the local moment $\mu$;
$J_{1b}$ (between parallel neighboring spins) can even change sign at large
values of $\mu$ \cite{FESE,Savrasov}. Such strong dependence of the exchange
parameters on the local moment clearly rules out the traditional Heisenberg
model description for ferropnictides. Moreover, the anisotropy of exchange
parameters breaks the symmetry of the spin Hamiltonian, which thereby becomes
unsuitable beyond the spin-wave regime. In particular, magnetic phase
transitions can not be described with such anisotropic Hamiltonian.

Another problem of the Heisenberg model is that it fails to describe the
energies of noncollinear structures connecting the degenerate AFM domains.
Specifically, in this model with interaction of any range the AFM domains
with ordering vectors $(\pi,0)$ and $(0,\pi)$ are connected by a continuous
degenerate set of noncollinear states (at least at the mean-field level). In
clear contradiction with this model, band structure calculations reveal the
presence of a rather high energy barrier \cite{FESE,Yaresko} between
$(\pi,0)$ and $(0,\pi)$ states. In addition, it is difficult to reconcile the
Heisenberg model with the fact that the domain walls between different AFM
domains (both twin and antiphase) are very thin \cite{DWs}.

Failure of the Heisenberg model in ferropnictides led to the belief that the
spin Hamiltonian approach has to be abandoned altogether in favor of a model
that explicitly takes into account fermionic degrees of freedom. Strong
anisotropy of the magnetic interaction in the stripe phase was argued to be
linked with orbital ordering \cite{Ferroorb}. In the model with a
non-interacting fermion sector added to the local moment subsystem, the
double exchange is anisotropic in the stripe phase \cite{DEx}. Although this
model may be able to reproduce the spin wave spectrum, it contradicts
transport measurements showing a rather moderate in-plane anisotropy of the
resistivity, whose sign is opposite to that expected in the double exchange
picture \cite{resist}. In addition, it is not well suited for the studies of
magnetic thermodynamics.

Thus, the Heisenberg model of any range clearly fails to describe the
magnetism of ferropnictides. It is therefore natural to request a more
refined theory which would (1) preserve the symmetry of the lattice, (2)
generate the correct spin wave spectrum and anisotropic linear response
properties in the stripe phase, (3) eliminate the spurious continuous ground
state degeneracy, (4) be suitable for the studies of magnetic thermodynamics
at finite temperatures, and (5) be consistent with electronic structure
calculations. It is also desirable for the model to be capable of reproducing
subtle features of magnetic thermodynamics observed in ferropnictides,
including the occurrence of both first- and second- order phase transitions
and the possibility of the nematic symmetry breaking. In this article we show
that \emph{all} these requirements can be met by including non-Heisenberg
interaction in the effective spin Hamiltonian. The resulting effective
Hamiltonian is found to explain numerous experimentally observed magnetic
properties.

\section{Results}

In the following, we justify the choice of the spin Hamiltonian, present an
alternative interpretation of neutron scattering experiments, and then
describe the magnetic thermodynamics in ferropnictides using the mean-field
approximation and classical Monte Carlo simulations.

\subsection*{Model spin Hamiltonian}

Our treatment is based on the following effective classical spin Hamiltonian:
\begin{equation}
H=\sum\limits_{i<j}\left[J_{ij}\mathbf{S}_{i}\mathbf{S}_{j}-\tilde
K_{ij}\left(\mathbf{S}_{i}\mathbf{S}_{j}\right)^{2}\right]  \label{A1}%
\end{equation}
The summation in (1) is taken over distinct pairs of lattice sites. The first
term is the conventional (isotropic) exchange interaction, and the second
term is the pairwise biquadratic interaction, which is the simplest and most
natural form of non-Heisenberg coupling. We found that a consistent
description of magnetism in ferropnictides is obtained using the
$J_{1}$-$K$-$J_{2}$-$J_{c}$ model, which includes the nearest and
next-nearest in-plane exchange parameters $J_1$ and $J_2$, the interplane
exchange parameter $J_c$, and the nearest-neighbor (in-plane) biquadratic
coupling $K=\tilde K S^2$.

Note that a weak effective biquadratic interaction appears in the
$J_{1}$-$J_{2}$ model through the ``order-from-disorder'' mechanism
\cite{Larkin}, and for pnictides it has been included in Ref.\ \cite{Fang}
assuming a very small interlayer coupling. Here we argue that a fairly strong
biquadratic interaction is already present in the microscopic spin
Hamiltonian in ferropnictides. Although magnetostructural coupling can also
effectively induce non-Heisenberg magnetic interactions, direct calculations
\cite{Xius} show that this distortion in the AFM phase has almost no effect
on magnetic interactions.

The presence of non-Heisenberg terms in the Hamiltonian radically changes the
interpretation of the LR exchange parameters $J_{ij}^{\mathrm{LR}}$, which
are defined as second derivatives of the total energy with respect to spin
rotations. Rotating the spins at sites $i$ and $j$ in the opposite directions
by an angle $\theta/2$, we find
\begin{equation}
J_{ij}^{\mathrm{LR}}\equiv-(\mathbf{S}_i\mathbf{S}_j)^{-1}\frac{\partial^{2}H}{\partial\theta^{2}}
=J_{ij}-2\tilde K_{ij}S^{2}\mathbf{e}_{i}\mathbf{e}_{j},
\end{equation}
where $\mathbf{e}_{i}=\mathbf{S}_{i}/S$. The contribution of the biquadratic
term depends on whether the spins in the given pair are parallel or
antiparallel. For the stripe ground state, we find $J_{1a}=J_{1}+2\tilde
KS^{2}$ and $J_{1b}=J_{1}-2\tilde KS^{2}$. Thus, the anisotropy of the
nearest-neighbor LR exchange parameter in the stripe phase is captured by the
biquadratic term in Hamiltonian (\ref{A1}).

\subsection*{Spin wave spectrum}

The spin wave spectrum of CaFe$_{2}$As$_{2}$ was measured using inelastic
neutron scattering in Refs.\ \cite{AMESneut,ORNLneut} and fitted to the
anisotropic $J_{1}$-$J_{2}$-$J_{c}$ Heisenberg model. Here we reinterpret
this fitting in terms of our \emph{isotropic} $J_{1}$-$K$-$J_{2}$-$J_{c}$
model, which produces an \emph{identical} spin wave spectrum. The fitted
value of $K$ is strongly affected by the inclusion of the zone-edge magnons
\cite{ORNLneut}. Table 1 shows the reinterpreted parameters of this fitting.

\begin{table}[htb]%
\begin{tabular}
[c]{|c|c|c|c||c|c|c|c|c|c|}\hline
$SJ_{1}$ & $SK$ & $SJ_{2}$ & $SJ_{c}$ & $J_{s}S^{2}$ & $J_{2}/J_{1}$ &
$J_{c}/J_{1}$ & $K/J_{s}$ & $T_{N}$, K & Order\\\hline
22 & 14 & 19 & 5.3 & 35 & 0.86 & 0.24 & 0.16 & 90 & I\\\hline
\end{tabular}
\caption{Parameters of the model (in meV) fitted to spin wave spectrum
\cite{ORNLneut}. Here $K=(J_{1a}-J_{1b})/4$, and $J_{s}=4J_{2}+2J_{c}$. The
parameter $J_{s}$ and $T_N$ are found assuming $S=0.4$.}%
\end{table}

\begin{figure}[tbh]
\includegraphics[width=0.45\textwidth,clip]{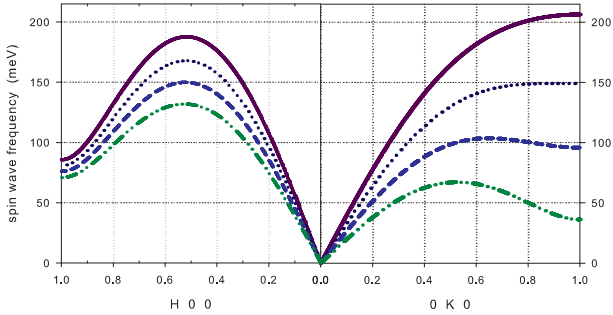}\caption{Effect of the
biquadratic coupling on the spin wave spectrum. The top curves correspond to
the model parameters taken from Table 1. To obtain the other curves, from top
to bottom, the ratio $K/J_{1}$ was reduced to $0.4$, $0.2$, and 0.}%
\label{SW}%
\end{figure}

Fig.\ \ref{SW} illustrates the effect of $K$ on the spin wave spectrum
described by the following dispersion relations:
\begin{equation}
E(\mathbf{q})=\sqrt{A_{\mathbf{q}}^{2}-B_{\mathbf{q}}^{2}}, \label{dispersion}%
\end{equation}
\noindent where
\begin{align}
A_{\mathbf{q}}  &  =2S[(J_{1}-2K)\cos(\pi k)+4K+J_{c}+2J_{2}]\label{a1}\\
B_{\mathbf{q}}  &  =2S[(J_{1}+2K)\cos(\pi h)+2J_{2}\cos(\pi h)\cos(\pi
k)\nonumber\\
&  +J_{c}\cos(\pi l)],\nonumber
\end{align}
where $\mathbf{q}$ is the reduced wave vector $(hkl)$. Starting from the
values taken from Ref.\ \cite{ORNLneut} (Table 1), the evolution of the
spectrum as $K$ is decreased to zero is shown. As noted in
Ref.\ \cite{ORNLneut}, the absence of a minimum at the zone edge along the
[010] direction signals a sign inversion of the parameter $J_{1b}$. Using spin
Hamiltonian (1), this feature is immediately reinterpreted as a signature of a
large value of biquadratic coupling $K>J_{1}/2$.

\subsection*{Mean-field phase diagram}

In the presence of biquadratic interaction the single-site problem in MFA
contains dipole and quadrupole effective fields with self-consistency
required for both the magnetization $m_{i}=\langle\mathbf{e}_{i}\rangle$ and
quadrupole order parameter $m_{ij}=\langle\mathbf{e}_{i}
\mathbf{e}_{j}\rangle -\delta_{ij}/3$. Thus, apart from an AFM (stripe) phase
with nonzero $m_{i}$, an anisotropic quadrupolar phase may appear with
$m_{i}=0$ but $m_{ij}\ne0$ \cite{Matveev}.

The MFA phase diagram is shown in the inset of Fig.\ \ref{MC}. The only
relevant phase up to $K/J_{1}=3$ is the AFM (stripe) phase. The most
important feature for our purposes is the tricritical point at
$K_{t}=5J_{s}/24$, where $J_{s}=4J_{2}+2J_{c}$. At $K=K_{t}$ the phase
transition to the paramagnetic phase changes from second order (at lower $K$)
to first order. At $K<K_{t}$ the transition temperature does not depend on
$K$ in MFA, while at $K>K_{t}$ there is a gradual crossover to a linear
dependence. The tricritical point falls well within he range of realistic
parameters for ferropnictides, and it is further significantly reduced by
fluctuations (see below). Thus, magnetic Hamiltonian (1) can yield a
first-order transition in ferropnictides without introducing any
magnetostructural coupling (as it was done in Refs.\ \cite{MS}). This is a
rather general feature of spin Hamiltonians with biquadratic coupling
\cite{Matveev,Nagaev}.

At $K\approx J_{s}$ there is a triple point where the quadrupolar phase
appears between the AFM and the paramagnetic phases. The quadrupolar phase is
always separated from the paramagnetic phase by a first-order transition, but
the AFM-to-quadrupolar transition changes from first to second order at a
slightly higher $K$. The triple point occurs beyond the reasonable range of
$K$ for ferropnictides.

\subsection*{Monte Carlo simulations}

\begin{figure}[tbh]
\includegraphics[width=0.45\textwidth]{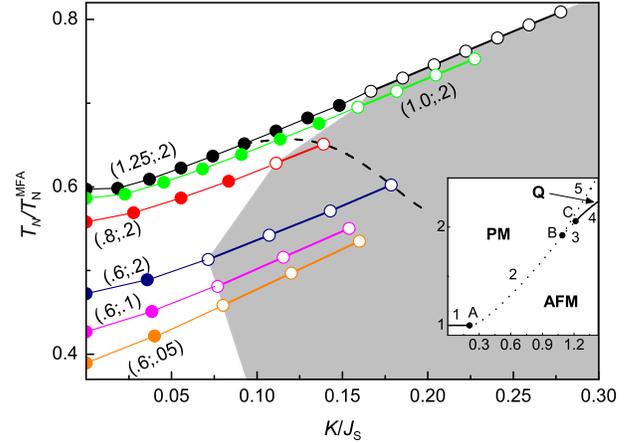}\caption{Transition temperature
and order of phase transition as a function of biquadratic interaction $K$
obtained by Monte Carlo simulations. Temperature is measured in units of
$T_N^{\textrm{MFA}}=J_{s}/3$, which is the second-order $T_{N}$ in MFA. Each
line is labeled by a set of two parameters ($J_{2}/J_{1}$; $J_{c}/J_{1}$).
The region of first-order transitions (empty symbols) is schematically
highlighted by shading and thicker lines. The dashed line shows the point of
inversion of $J_{1b}$. The inset shows the mean-field phase diagram in a
larger area of the same parameters. Solid lines 1 and 4 (dashed lines 2, 3
and 5) denote second-order (first-order) first transition. A and C are
tricritical points; B is the triple point. Q denotes the quadrupolar phase.}%
\label{MC}%
\end{figure}

The main results are presented in Fig.\ \ref{MC}. We considered a few values
of the $J_{2}/J_{1}$ ratio from 0.6 to 1.25 at fixed $J_{c}/J_{1}$=$0.2$, and
a few values of $J_{c}/J_{1}$ at $J_{2}/J_{1}=0.6$. Note that
$J_{2}/J_{1}=0.5$ is the stability limit of the stripe phase at $T=0$, and
one can expect particularly strong fluctuations in the vicinity of this
point. The fluctuations are also enhanced at small $J_{c}/J_{1}$, where
two-dimensional behavior sets in.

As expected, the transition temperature is significantly reduced compared to
MFA, and this reduction increases as $J_{2}/J_{1}$ approaches 0.5 and as
$J_{c}/J_{1}$ decreases. Interestingly, the ratio $K_{t}/J_{s}$ demonstrates a
similar trend. The critical point for the parameters corresponding to
CaFe$_{2}$As$_{2}$ is estimated at 90 K, which, considering the simplicity of
the model, is in reasonable agreement with the experimental range of 140-170 K
for 122 compounds. Further, this transition is of first order for these
parameters, in agreement with experiment.

\begin{figure}[htb]
\includegraphics[width=0.4\textwidth]{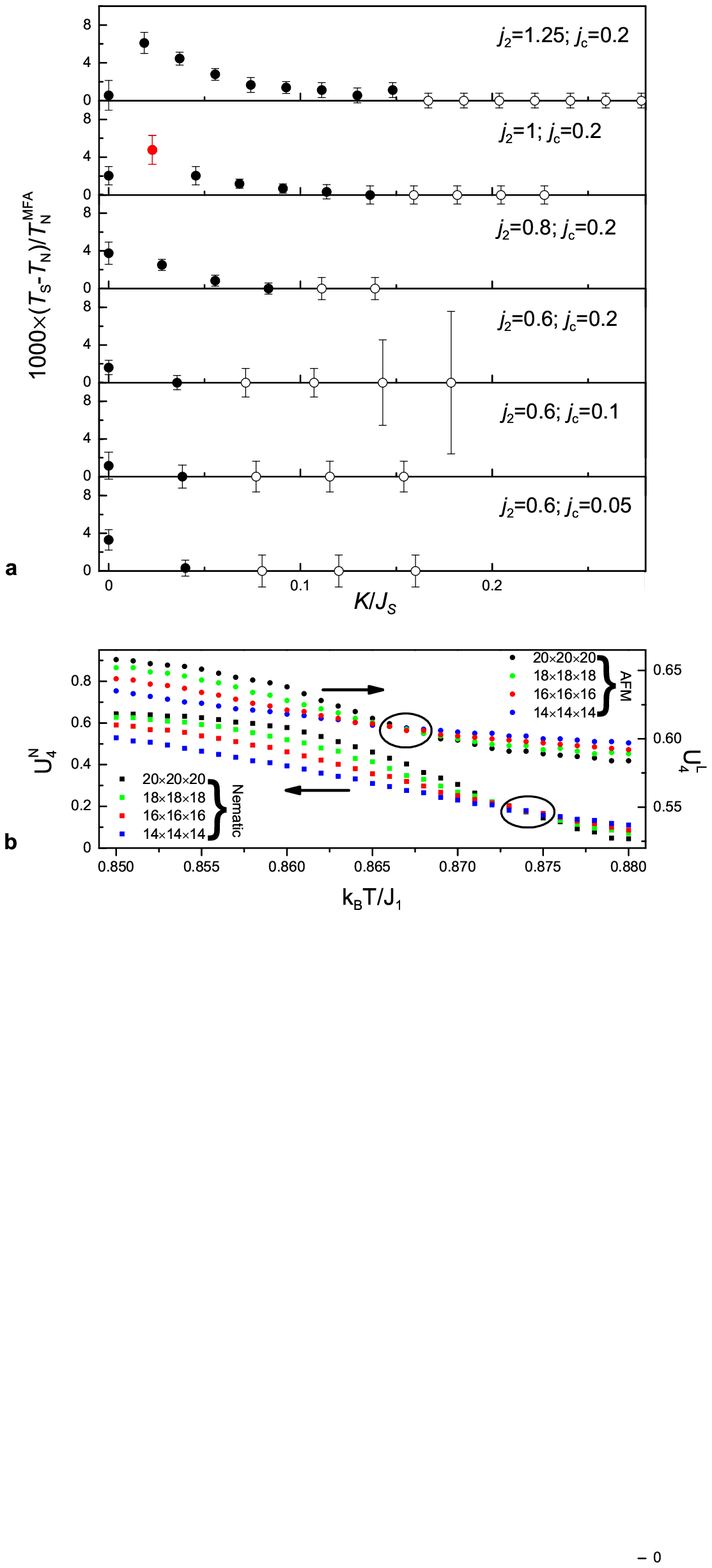} \caption{(\textbf{a}) Relative separation of the
nematic ($T_s$) and AFM ($T_N$) transition temperatures as a function of the
biquadratic interaction parameter $K$ for different values of
$j_{2}=J_{2}/J_{1}$ and $j_{c}=J_{c}/J_{1}$. Open (filled) symbols correspond
to the first-order (second-order) phase transition. (\textbf{b}) Example of
the separation of $T_s$ and $T_N$ in the finite-size scaling behavior of the
respective fourth-order cumulants $U^N_4$ and $U^L_4$; the parameters are
for the point shown in red in panel \textbf{a}.}%
\label{MC2}%
\end{figure}

Experimentally, a second-order AFM phase transition in ferropnictides is
typically preceded by an orthorhombic structural transition. Between these
transitions the so-called nematic phase occurs \cite{Johnston}.
Theoretically, the nematic phase is known to occur at $J_{c}=0$ \cite{Larkin}
and was predicted to occur at $J_{c}\ll J_{1}$ as well \cite{Fang}. It is not
clear whether this phase can survive in the presence of substantial
interplane and biquadratic interactions. It is rather difficult to establish
the separation of the nematic and AFM phase transitions in Monte Carlo
simulations, because the AFM correlation length becomes very large whenever
this separation becomes large, and therefore a spurious
``quasi-long-range-order'' is generated by finite-size effects \cite{Kapik}.
However, in a certain parameter range we have observed a clear signature of
the nematic phase in the finite-size behavior of the fourth-order cumulants
of the nematic and AFM order parameters (their definitions are given in the
\emph{Methods} section). Fig.\ \ref{MC2}a shows the separation between the
nematic and AFM phase transitions determined from the crossing points of the
fourth-order cumulants for these order parameters calculated for different
$D$ (an example is shown in Fig.\ \ref{MC2}b). The separation determined in
this way appears to increase at larger $J_{2}/J_{1}$, smaller $K$, and
smaller $J_{c}$. It never occurs in the range of parameters where the phase
transition is of the first order.

\section{Discussion}

Since the $K/J$ ratios are proportional to $S^{2}$, the AFM phase transition
can change from first to second order if $S$ is reduced. Such behavior was
indeed observed under electron doping which reduces the local moments.
Indeed, while the phase transition in updoped 122 compounds
($\mu\sim0.8-0.9\mu_{B}$) appears to be first-order (most clearly in
CaFe$_2$As$_2$ \cite{1order}), it changes to second order under electron
doping \cite{BAFECO,Fernandez}. While the decrease of amplitude of the local
moments in doped systems has not been measured, the critical temperature of
magnetic phase transition and the order parameter are clearly decreased with
doping. The band structure studies further support this trend. The
experimental situation with the local moment and order of the phase
transition in LaFeAsO and other 1111 compounds is not yet fully resolved, but
recent measurements suggest that they may be closer to 122 compounds than
previously thought \cite{Braden}.

In view of the close proximity of the nematic and AFM phase transitions and
the corresponding limitations of the finite-size scaling procedure, the
signature of the separation of these transitions, while statistically robust,
can not be viewed as the final proof. Still, it is remarkable that the
separation appears to survive in an essentially three-dimensional case. In
addition, contrary to Ref.\ \onlinecite{Fang}, the biquadratic interaction is
present on the level of microscopic Hamiltonian and is not small. Curiously,
Fig.\ 3 suggests that at certain conditions the separation can go through a
maximum as a function of $K$. The origin of this behavior remains to be
understood.

Comparison of the effective spin Hamiltonian fitted to experiment (Table 1)
with the results of first-principles calculations leads to an interesting
observation. The experimental exchange parameters in Table 1 agree quite well
with LR calculations \cite{Savrasov,FESE}, except that the calculated value
of $K$ (which can be deduced either from the anisotropy of $J_1$ or from the
height of the barrier between the two stripe states) reaches its experimental
magnitude only for larger values of the local moment of order 1.4--1.8
$\mu_B$. First-principles analysis of the dynamic susceptibility \cite{Xius}
also shows good agreement with experimental spin wave spectrum if the local
moment is of that order. On the other hand, scaling of the spin moment up by
a factor of two would also scale up the predicted $T_N$ (Table 1) by a factor
of two and bring it in almost quantitative agreement with experiment. This
observation seems to support the notion that the (mean-field) local moments
in ferropnictides are roughly twice their observed values, which may be
screened by dynamical fluctuations.

We have shown that a consistent model of magnetism in ferropnictides is
obtained by introducing biquadratic interaction in the model spin
Hamiltonian. This model satisfies all reasonable requirements, reproduces
many experimental observations, and is supported by electronic structure
calculations. As a word of caution, the success of the localized model spin
Hamiltonian should not be taken to rule out itinerant effects. On the
contrary, the effective spin model should be viewed as an appropriate mapping
of the complicated itinerant interactions, which include complicated
rearrangements of the Fermi surface and density matrices induced by magnetic
ordering. This interpretation is supported by the fact that a large
biquadratic term appears at relatively low values of the local moment. We
find it appropriate to speculate that strong non-Heisenberg coupling in the
effective spin model, being a consequence of strong coupling between the
itinerant electrons and spin fluctuations, explicitly suggests spin
fluctuations as a possible pairing mechanism in ferropnictides.

\section{Methods}

We have studied tetragonal lattices $D\times D\times D$ with periodic
boundary conditions, where $D=14$, 16, 18, and 20. Looping over the lattice
sites, a new random spin direction for the given spin is tried and accepted
or rejected using the Metropolis prescription. The averages are accumulated
during a sufficiently long measurement run after an initial equilibration
run. The lengths of these runs were adjusted in order to obtain sufficient
accuracy. The AFM order parameter $L$ is defined as
$L^{2}=L_{x}^{2}+L_{y}^{2}$, where $L_{x}$ and $L_{y}$ are the staggered
order parameters with the $x$ and $y$ staggering direction, respectively. The
nematic order parameter is defined as $N=\left\vert
\frac{1}{2}\sum_{i}\sum_{nn}\eta_{ij}\mathbf{S}_{i}\cdot\mathbf{S}_{j}\right\vert$,
where the sum is over the four in-plane nearest neighbors, and $\eta_{ij}$ is
equal to $\pm1$ with opposite signs taken for the two perpendicular
directions. The order of the phase transition was determined by analyzing the
behavior of the fourth-order energy cumulant \cite{Challa}. If this cumulant
converges toward the value $2/3$ for all temperatures with increasing $D$,
the transition is of second order. On the other hand, if the energy cumulant
develops a minimum at some temperature, which becomes sharper with increasing
$D$, the transition is of first order \cite{Challa}. If the transition was
found to be of second order, the critical temperatures for the AFM and
nematic transitions were determined from the finite-size scaling behavior of
the fourth-order cumulant of the respective order parameter \cite{Binder}.
For first order transitions, the transition temperatures for AFM and nematic
phase transition were found from the peaks of the corresponding
susceptibilities.

\section*{Acknowledgments}

We are grateful to S. Bud'ko for fruitful discussions, and to I. I. Mazin for
critical reading of the manuscript. Work at UNL was supported by NSF
DMR-1005642 and EPS-1010674. Work at Ames Laboratory was supported by
Department of Energy-Basic Energy Sciences, under Contract No.
DE-AC02-07CH11358. K.\ B.\ is a Cottrell Scholar of Research Corporation.

\end{document}